\documentclass[prl,twocolumn,amsmath,amssymb]{revtex4-1}

\usepackage{amsmath,latexsym}
\usepackage{graphicx}
\usepackage{amssymb}
\usepackage{amsfonts}
\usepackage{amsbsy}
\usepackage{natbib}
\usepackage{comment}
\usepackage{color}
\usepackage{float}

\usepackage[colorlinks=true, urlcolor=blue, citecolor=blue, linkcolor=blue]{hyperref}

\renewcommand{\vec}[1]{\mbox{\boldmath$\mathrm{#1}$}}
\let\sb=_ \catcode`\_=\active \def_#1{\ensuremath \sb{\rm#1}}

\renewcommand{\vec}[1]{\mbox{\boldmath$\mathrm{#1}$}}
\newcommand{\ii}{\mathrm{i}}
\newcommand{\ee}{\mathrm{e}}

\begin{document}

\title{$\mathcal{PT}$-Symmetric Magnon Lasing and Anti-Lasing}

\author{Xi-guang Wang$^{1}$, Tian-xiang Lu$^{2}$, Guang-hua Guo$^{1}$, Jamal Berakdar$^{3*}$, Hui Jing$^{4*}$}

\address{$^1$ School of Physics, Central South University, Changsha 410083, China \\
	$^2$ College of Physics and Electronic Information, Gannan Normal University, Ganzhou 341000, China \\
	$^3$ Institut f\"ur Physik, Martin-Luther Universit\"at Halle-Wittenberg, 06099 Halle/Saale, Germany \\
	$^4$ Key Laboratory of Low-Dimensional Quantum Structures and Quantum Control of Ministry of Education, Department of Physics and Synergetic Innovation Center for Quantum Effects and Applications, Hunan Normal University, Changsha 410081, China \\
	$^*$ email: jamal.berakdar@physik.uni-halle.de; jinghui@hunnu.edu.cn}

\date{\today}

\begin{abstract}
	A mechanism for electrically tunable $\mathcal{PT}$-symmetric magnonic lasing and anti-lasing is proposed along with a device consisting of a current-biased region in a magnetically ordered planar  waveguide. Within the bias area, several heavy-metal wires carrying dc charge current are periodically attached to the waveguide and exert  so spatially  periodic  spin–orbit torques, producing current-controllable modulated magnon gain and loss. It is demonstrated  that this decorated waveguide  can emit a strong, single frequency magnon mode at the Bragg point (lasing) and also   absorb at the same frequency  phase-matched incoming coherent magnons  (anti-lasing). The underlying physics is captured by an analytical model   and validated   with full material and device-specific numerical  simulations. The magnonic laser absorber response is tunable via the current density in the wires, the extent of the biased region, and the intrinsic damping, enabling the control of  lasing frequency and emission power. The structure is shown to amplify thermal magnons, offering a route to low-noise on-chip microwave sources. The concept is compatible with planar waveguides, ring geometries, and antiferromagnets. The results establish an experimentally realistic platform where a single element functions simultaneously as both magnon laser and  absorber, opening opportunities for reconfigurable non-Hermitian magnonics and integrated magnon signal processing.
\end{abstract}

\maketitle

Magnons have been demonstrated to function as carriers of  information  and to enable  data processing with several advantages related to their short wavelength, controlability  with various external stimuli, and ready integration in (spin)electronic circuits  \cite{Chumak2015,50007095,Pirro2021,Kruglyak2010,Serga2010,Vogt2014,Chumak2014ncomms5700,Tsoi2000nature}.  A prerequisite for magnonic functionality is the ability to generate and amplify coherent magnons in magnetically active nanostructures.
 To realize a magnon laser emitting coherent magnons, numerous studies utilized spin torque oscillators converting a dc current into  coherent magnons \cite{Tsoi2000nature,Kajiwara2010nature,Merbouche2024naturecommun,Collet2016,PhysRevApplied13044050,Demidov2010naturemater,Duan2014naturecommun, Demidov2017physrep}. In spin torque oscillators, a dc current injects  a spin current into the magnetic layer exerting a anti-damping torque (gain) on the local magnetization oscillations to  compensate the natural oscillation loss and to amplify the magnon amplitude \cite{Kajiwara2010nature,Merbouche2024naturecommun,doi101126science1105722,Collet2016,Hoffmann2013,Garello2013,PhysRevLett110147601, RevModPhys871213, RevModPhys91035004, Haidar2019, Fulara2019, Houssameddine2007, Kaka2005}. For such scheme of  electrical magnon generators based  on gain, the magnon coherence and single-mode purity are usually limited by multi-mode excitations. Here,  a qualitatively new concept is presented for purely amplifying  single magnon modes of interest in a narrow frequency range: This is achieved  by proposing   a setup with a  spatially inhomogeneous spin-orbit torques (SOTs) that cause  periodic gain and loss of magnetization oscillations. The gain and loss are balanced and the system is shown to exhibit 
  a parity-time($\mathcal{PT}$) symmetry which has been exploited in  optics \cite{Feng2017NaturePhoton,PhysRevLett.103.093902, Ruter2010naturephys}, acoustics \cite{PhysRevX4031042, 101093nsrnwy011, Fleury2015naturecommun}, electronics \cite{PhysRevA84040101}, and now magnonics \cite{Wangxinc2020, PhysRevApplied034050, PhysRevApplied.18.024073, Lee2015, Liu2019, PhysRevApplied18014003, PhysRevLett.125.147202, Wittrock2024}. In addition, it is documented  that involving $\mathcal{PT}$-symmetry in laser concepts  leads to better performance. For example, one can enhance desired mode's output while suppressing others \cite{doi101126science1258480,PhysRevLett113053604}. An interesting finding in this area is that, with fine tuning in gain/loss, the $\mathcal{PT}$-symmetry-assisted  laser can both emit and absorb  coherent waves, meaning the ability for lasing and anti-lasing \cite{PhysRevA82031801,PhysRevA110033504,Wong2016naturephoton}. Recently, the cavity optomagnonic system was exploited to combine $\mathcal{PT}$-symmetry laser and magnon excitation in a yttrium iron garnet (YIG) sphere \cite{PhysRevA105053705, Wang24optexpress, bnynmbwv2025}.

\begin{figure}[htbp]
	\includegraphics[width=0.5\textwidth]{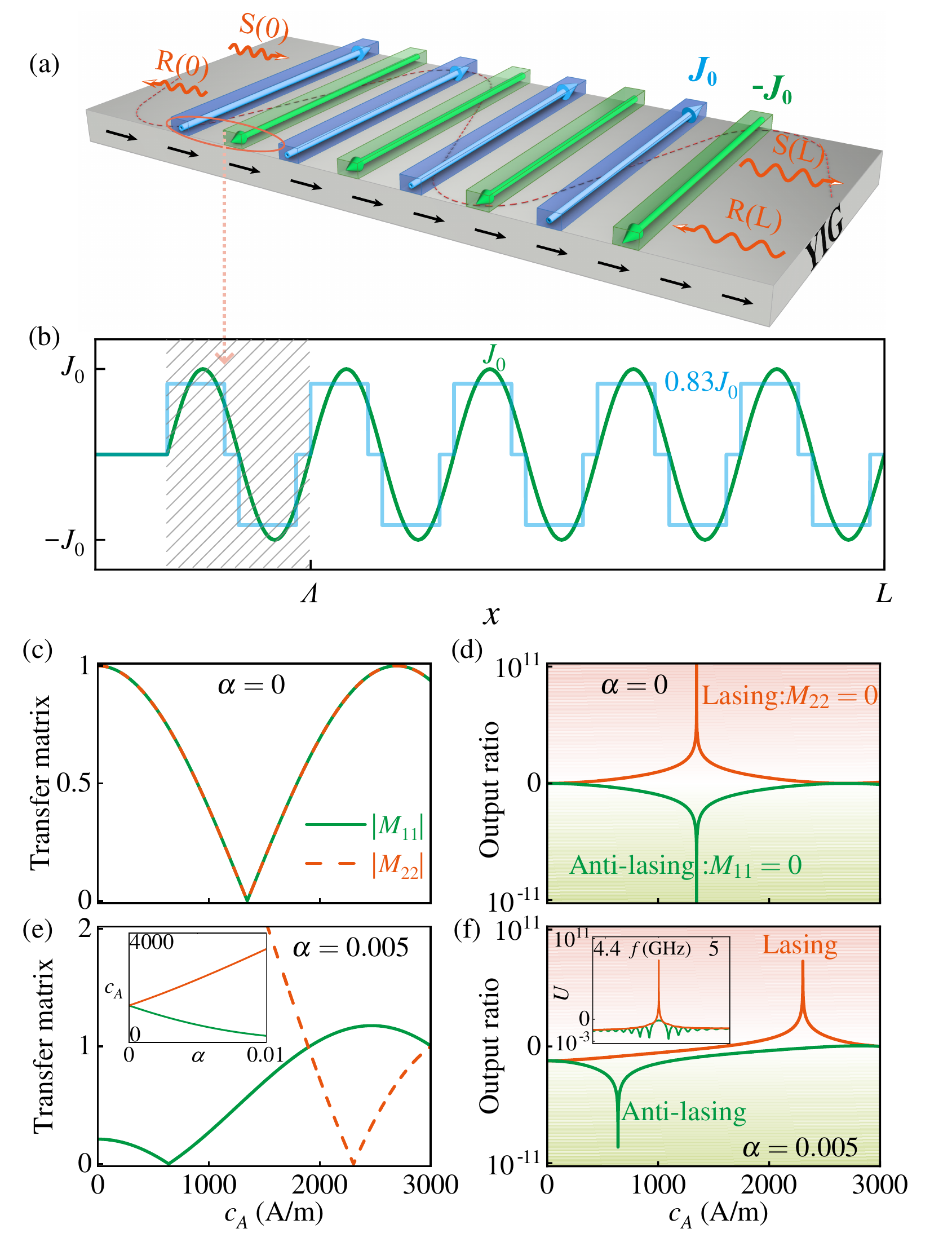}
	\caption{\label{model} (a)  Schematic of a $\mathcal{PT}$-symmetry-based magnonic laser absorber. A number of electrically separated heavy metal stripes carrying opposite charge current densities  $J_c(x)$ (flowing in $y$ direction) are attached to a magnonic planar waveguide, generating alternating gain and loss of magnetic oscillations. (b) Spatial profiles of the sine and step functions of current density $J_c(x)$ adopted in the model. The period length $\Lambda = 100$ nm, and the total SOT region length is $ L = 5000$ nm. For (c-d) $\alpha = 0$ and (e-f) $\alpha = 0.005$, transfer matrix elements ($M_{11}$, $M_{22}$) and output ratio $U$ (logarithmic scale) dependence on the electric current amplitude $ c_{A} $ at the Bragg point $\Re[k_x] = k_p$. The inset of (e) shows the $\alpha$ dependent positions of $M_{11} \approx 0$ (anti-lasing condition) and $M_{22} \approx 0 $ (lasing condition). The inset of (f) is the frequency $f$ dependent $U$ under the lasing and anti-lasing conditions. The lasing case responds to the single input excitation [$S(0) \ne 0$, $R(L) = 0$], and anti-lasing case means  magnon excitation under two coherent inputs with $ R(L)/S(0) = M_{21} $.  Results follow from Eq. (\ref{transfer}). }
\end{figure}

The current proposal for $\mathcal{PT}$-symmetric magnonic lasing and anti-lasing is realizable  in a single magnetic planar waveguide without the need for optical pumping (c.f. Fig. \ref{model}). The system leverages the unique magnonic features such as swift manipulation via dc current or magnetic field and nonlinear magnon interaction. By reaching a critical dc current density, the region with gain and loss  emit only single mode magnons at the Bragg point, and nonlinear effect only slightly shift the magnon frequency around the Bragg point without amplifying other modes. On the other hand, the structure can absorb the magnons at the same frequency under two coherent inputs, achieving magnon anti-laser. Such functionality is in line  with the growing interest in magnetically tunable absorption \cite{Qian2025naturecommun, PhysRevB102054429, Rao2021naturecommun}.   Based on  an analytical model necessary conditions are presented in this work for single magnon mode laser, absorber and their manipulation mechanisms, and the influence of intrinsic Gilbert damping is analyzed.  Full-fledged  numerical  simulations confirm the analytical prediction and enable  application of the scheme to various structures include magnetic rings and antiferromagnet in a material and device-specific manner. The results are potentially useful for developing new highly tunable magnonic devices merging single magnon mode oscillator and absorber.

To generate  $\mathcal{PT}$-symmetric structure in a single magnonic ferromagnetic palnar waveguide (along $x$ axis), one attaches  charge-current carrying heavy-metal stripes  which amounts to spatially varying electrical current density $J_c(x)$ flowing along the $y$ axis, as illustrated  in Fig. \ref{model}. $J_c(x)$  is perceived in the waveguide as  a SOT $\vec{T} = \gamma c_J \vec{m} \times \vec{e}_x \times \vec{m} $ with the strength $c_{J} = \theta_{\mathrm{SH}} \frac{\hbar J_c}{2\mu_0 e d M_{\mathrm{s}}}$. Here, $M_s$ is the saturation magnetization,  $\theta_{SH}$ is the spin-Hall angle, $d$ is the thickness of magnetic layer, and $\gamma$ is the gyromagnetic ratio. The magnonic modes  of the waveguide  are well described  by the Landau-Lifshitz-Gilbert (LLG) equation \cite{Hoffmann2013,Garello2013, RevModPhys871213},
\begin{equation}
	\displaystyle \frac{\partial \vec{m}}{\partial t} = - \gamma \vec{m} \times \vec{H}_{\rm{eff}}+  \alpha \vec{m} \times \frac{\partial \vec{m}}{\partial t} + \vec{T},
	\label{LLG}
\end{equation}
where $\vec{m}$ is the unit vector of local magnetization, and $\alpha$ is the intrinsic Gilbert damping of the magnetic waveguide. The effective field $H_{eff} = \frac{2 A_{ex}}{\mu_0 M_s} \nabla^2 \vec{m} + H_{0} \vec{e}_x$ consists of the internal exchange field (with the exchange constant $A_{ex}$), and the external magnetic field $H_0$. Under the effective field, the equilibrium magnetization is in the $\vec{e}_x$ direction. In numerical calculations, we adopt following parameters of YIG: $M_s = 1.4 \times 10^5$ A/m, $A_{ex} = 3 \times 10^{-12}$ J/m, $H_0 = 1 \times 10^5$, and $\alpha = 0.005$.

We start with a $\mathcal{PT}$-symmetric magnonic laser absorber structure generated by the periodic sine function  $c_J(x) = c_A \sin (2 k_p x) $  with period length $\Lambda = \pi/k_p$, as shown in Fig. \ref{model}(b). The periodic region is located within $ 0 \le x \le L$, where $ L = n \Lambda$ is the integer multiple of the period length. To setup an analytical model for the magnon scattering in the periodic structure, we introduce small transverse deviation $ \vec{m}_{\mathrm{s}} \ee^{-\ii \omega t} $, $\vec{m}_{\mathrm{s}} = (0, \delta m_y, \delta m_z)$ from the equilibrium $\vec{m}_0 = \vec{e}_x$. Substituting the magnon function and defining $ \psi = \delta m_y - \ii \delta m_z $, the linearized LLG equation (\ref{LLG}) is reformulated  in the Helmholtz equation,
 \begin{equation}
	\begin{aligned}
		\displaystyle 
		\label{helm} \psi''(x)+\left[\frac{\omega-\omega_H}{\omega_k} + \ii \frac{\omega_J(x)}{\omega_k}\right]\psi(x)=0.
	\end{aligned}
\end{equation} 
Following notions are introduced: $\omega_k = 2 (1 - \ii\alpha) \gamma A_{\mathrm{ex}} /[\mu_0 M_{\mathrm{s}} (1+\alpha^2)]$, $\omega_{J}(x) = (1-\ii\alpha) \gamma c_{J}(x)/(1+\alpha^2)$, and $\omega_H = (1 - \ii\alpha) \gamma H_0/(1+\alpha^2)$. Around the nodes $x = n \Lambda$ of the periodic potential, the electrical current term $c_J(x)$ induced gain and loss is asymmetric. Thus, the coefficient $ O_{\mathrm{H}} = (\omega-\omega_H)/\omega_k + \ii \omega_J/\omega_k $ in Eq. (\ref{helm}) satisfies the $\mathcal{PT}$-symmetry condition $ (O^l_{\mathrm{H}})^* = O^r_{\mathrm{H}} $ by neglecting $ \alpha $, where $ O^l_{\mathrm{H}} $ and $ O^r_{\mathrm{H}} $ are the coefficients in the left and right unit of the integer nodes.

In the periodic structure, we assume  forward ($+x$ direction) and backward ($-x$ direction) propagating magnons to have the form  $ \psi(x) = S(x)\ee^{\ii k_p x} + R(x)\ee^{-\ii k_p x}  $. Substituting this expression into Eq. (\ref{helm}), one infers  the  equation governing the magnons  near the Bragg point ($k_x \approx k_p$  ),
 \begin{equation}
	\begin{aligned}
		\displaystyle 
		\label{coupled} -\ii R'(x) + \delta_k R(x) &= \xi S(x),\\
		\ii S'(x) +  \delta_k S(x) &= -\xi R(x).
	\end{aligned}
\end{equation} 
Solving the above equation, we extract the magnon solution at $x = L$ in the form of $(S(L), R(L))^T = \hat{M} (S(0), R(0))^T$, and the transfer matrix $\hat{M}$ is,
 \begin{equation}
 	\begin{small}
	\begin{aligned}
		\displaystyle 
		\label{transfer} \hat{M} = \begin{pmatrix}
				\cos(\beta L) + \frac{\ii \delta_k}{\beta}\sin(\beta L) &
				\frac{\ii \xi}{\beta}\sin(\beta L) \\
				\frac{\ii \xi }{\beta} \sin(\beta L) & 
				\cos(\beta L) - \frac{\ii \delta_k}{\beta} \sin(\beta L)
			\end{pmatrix}.
 \end{aligned}
 \end{small}
  \end{equation} 
Here, $\omega_{JA}(x) = (1-\ii\alpha) \gamma c_{A}(x)/(1+\alpha^2)$, $\xi = \omega_{JA}/{4\omega_k k_p}$, $\delta_k = k_x - k_p$, $\beta = \sqrt{\delta_k^2 + \xi^2}$, and the magnon wavevector $k_x =  \sqrt{(\omega - \omega_H)/\omega_k}$. From the transfer matrix element, the left/right ($l/r$) transmission and reflection coefficients are determined as $t_l = t_r = 1/M_{22}$, $ r_l = -M_{21}/M_{22} $ and $ r_r = M_{12}/M_{22}  $, where the coefficients correspond to $t_l = S(L)/S(0)$, $t_r = R(0)/R(L)$, $r_l = R(0)/S(0) $ and $r_r = S(L)/R(L) $.

 \begin{figure}[tbp]
	\includegraphics[width=0.5\textwidth]{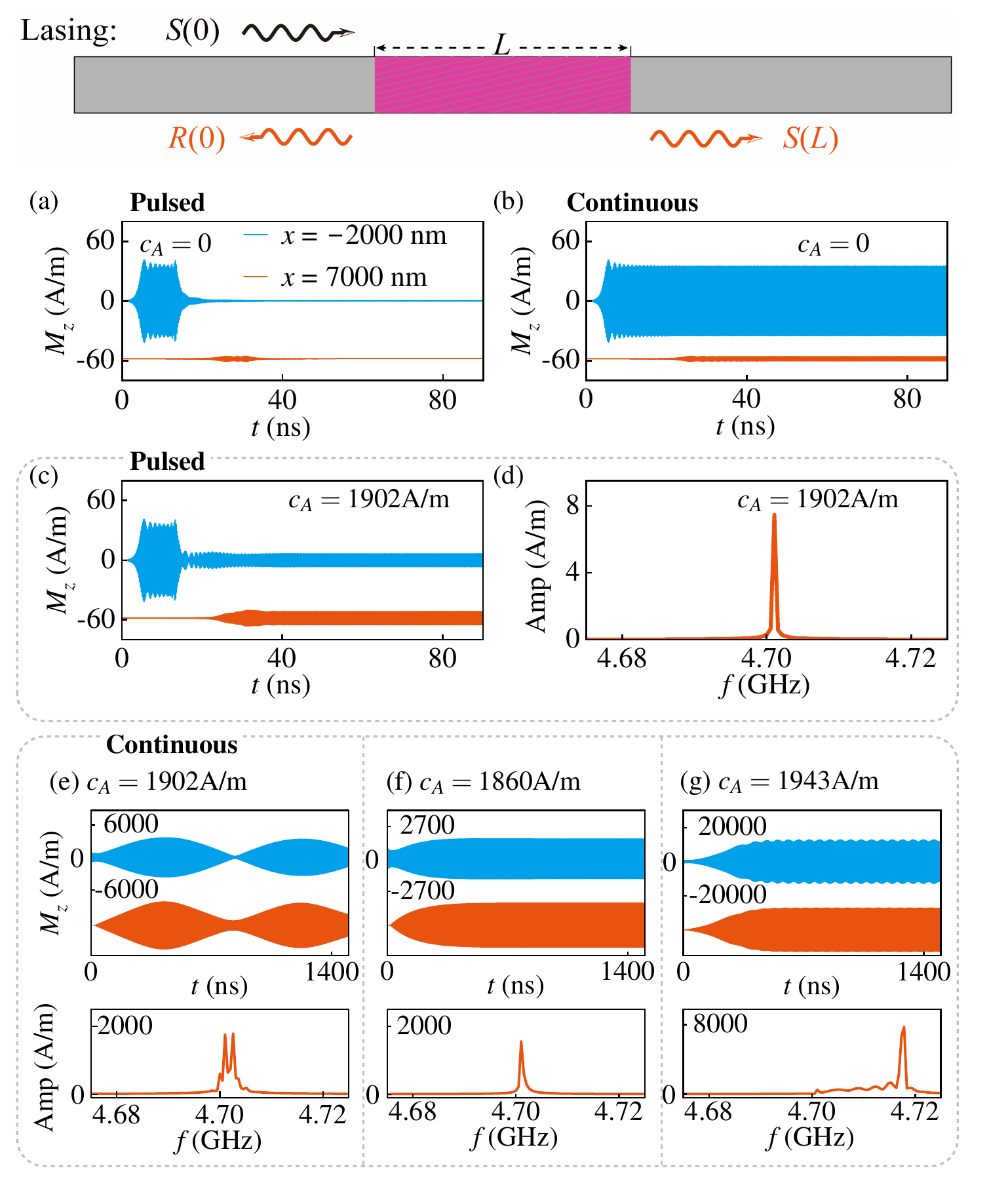}
	\caption{\label{laser} Top panel is the schematic for magnon lasing with single input $S(0) \ne 0$. The magnon outputs are detected as time-dependent $M_z(t)$ at left side $x = -2000$ nm [blue curve, related to $R(0)$] and right side $x = 7000$ nm [red curve, related to $S(L)$]. The red curves are in the same range with blue curves, which are intentionally offset for clarity. Under different $c_A$,  magnon outputs are excited by a microwave field $ \vec{h} = h_0 \sin(2 \pi f t) \vec{z} $ lasting 10 ns (a,c) and continuously (b,e, f, g), where frequency $f = 4.701 $ GHz and amplitude $h_0 = 1000$ A/m is localized at left side $x = -4000 $ nm. Frequency spectra are obtained from the laser induced long lasting oscillation.  These results are obtained from micromagnetic simulations with Eq. (\ref{LLG}), where the laser region realized by step-type function is limited in $ 0 \le x \le 5000$ nm with period $\Lambda = 100$ nm.}
\end{figure}

To achieve a magnonic laser with strong outputs [$R(0), S(L)$] under small inputs [$R(L), S(0)$], the element $M_{22} = 0$ is required. Meanwhile, by neglecting $ \alpha $, the $\mathcal{PT}$-symmetric scattering potential brings $M_{11} = M^*_{22} = 0$. Assuming two coherent magnon inputs with amplitude and phase defined by $R(L) = M_{21} S(0) $, the output solution becomes $R(0) = S(L) = 0$, representing a perfect absorber without any reflection, i.e. anti-lasing. Such amplification and absorption properties make the scattering potential behaving as both a magnonic laser and perfect absorber. Our numerical results in Fig. \ref{model}(c) validate these features. The lasing and anti-lasing conditions for $M_{11,22} = 0$ only can be reached at the Bragg point $\Re[k_x] = k_p$ when magnon frequency $f = \omega/(2\pi) = 4.701$ GHz and electric current amplitude $ c_{A} = 1346 $ A/m. In Fig. \ref{model}(d). we also calculate the output ratio $U$ of total outgoing magnon intensity $[R^2(0) + S^2(L)]$ to the incoming magnon intensity $[R^2(L) + S^2(0)]$, $U = [R^2(0) + S^2(L)]/[R^2(L) + S^2(0)]$. $ U > 1 $ represents the magnon amplification, and $U = 0$ means perfect absorption. For the single input [$S(0) \ne 0$, $R(L) = 0$], the output $ U $ takes very large value as the laser condition is approached. Injecting the second coherent signal $R(L) = M_{21} S(0)$, the behavior of $ U $ is obviously changed to perfect absorption at the same $c_A$ with a dip of $U \rightarrow 0$. 

In realistic case, the influence of finite Gilbert damping must be included. See Figs. \ref{model}(e-f) with $ \alpha = 0.005 $, the damping induced additional magnonic loss increases the amplitude $c_A$ of the lasing condition $M_{22} \rightarrow 0$, while anti-lasing condition $M_{11} \rightarrow 0$ related $c_A$ becomes smaller. The separation between the lasing and anti-lasing conditions increases with $\alpha$, as proved by the inset of Fig. \ref{model}(e). Still, the output ratio $U \gg 1$ reaches large value at the lasing condition $ c_{A} = 2305 $ A/m under single input [red curve in Fig. \ref{model}(f)]. The frequency dependent $U$ further proves that the laser is limited at the Bragg point with $f = 4.701$ GHz, see the inset of Fig. \ref{model}(f). For the anti-lasing at $c_A = 638$ A/m, a very small dip in $U$ is realized by two coherent injected inputs $R(L) = M_{21} S(0) $. Noteworthy, although the anti-lasing condition is not satisfied at the lasing $c_{A}$, we find the ratio $U = 0.67$ is still a small value under two inputs $R(L) = M_{21} S(0) $. Such coherent injections can be realized by combining magnon power divider and delay line/phase shifter. Thus, depending on the injected magnon inputs and the current density $c_A$, the periodic structure can either amplify or absorb the input magnons.

We need to note that, the above theoretical model for coupled modes of $S(x) \ee^{i k_p x}$ and $R(x) \ee^{-i k_p x}$, does not include the time variation component. The laser induced magnon amplification process requires a long enough duration. For describing the time effect, we substitute $ [S(x,t)\ee^{\ii k_p x} + R(x,t)\ee^{-\ii k_p x}]e^{-\ii \omega t} $ into the Eq. (\ref{LLG}). Combining Eq. (\ref{transfer}), we assume the solution profile follows $[S(t) + R(t)] \ee^{i \beta x}$, and the magnon attenuation time is extracted from the real component of $\tau = 2\ii\omega_{k}\delta_k k_p + \sqrt{\omega_{JA}^2/4 - 4 \omega_{k}^2 \beta^2 k_p^2}$. Using the value of $\beta$ at the laser condition with the critical $c_A$, the real component $\Re[\tau] = 0$ indicates unattenuated magnons, and above the laser critical value there are positive $\Re[\tau] > 0$ and amplifying magnons. Such features indicate that, instead of requiring a specific $c_A$, the predicated magnon laser mechanism always works for $c_A$ above the critical value. The conclusion is testified by following full wave numerical simulation with Eq. (\ref{LLG}).

\begin{figure}[tbp]
	\includegraphics[width=0.5\textwidth]{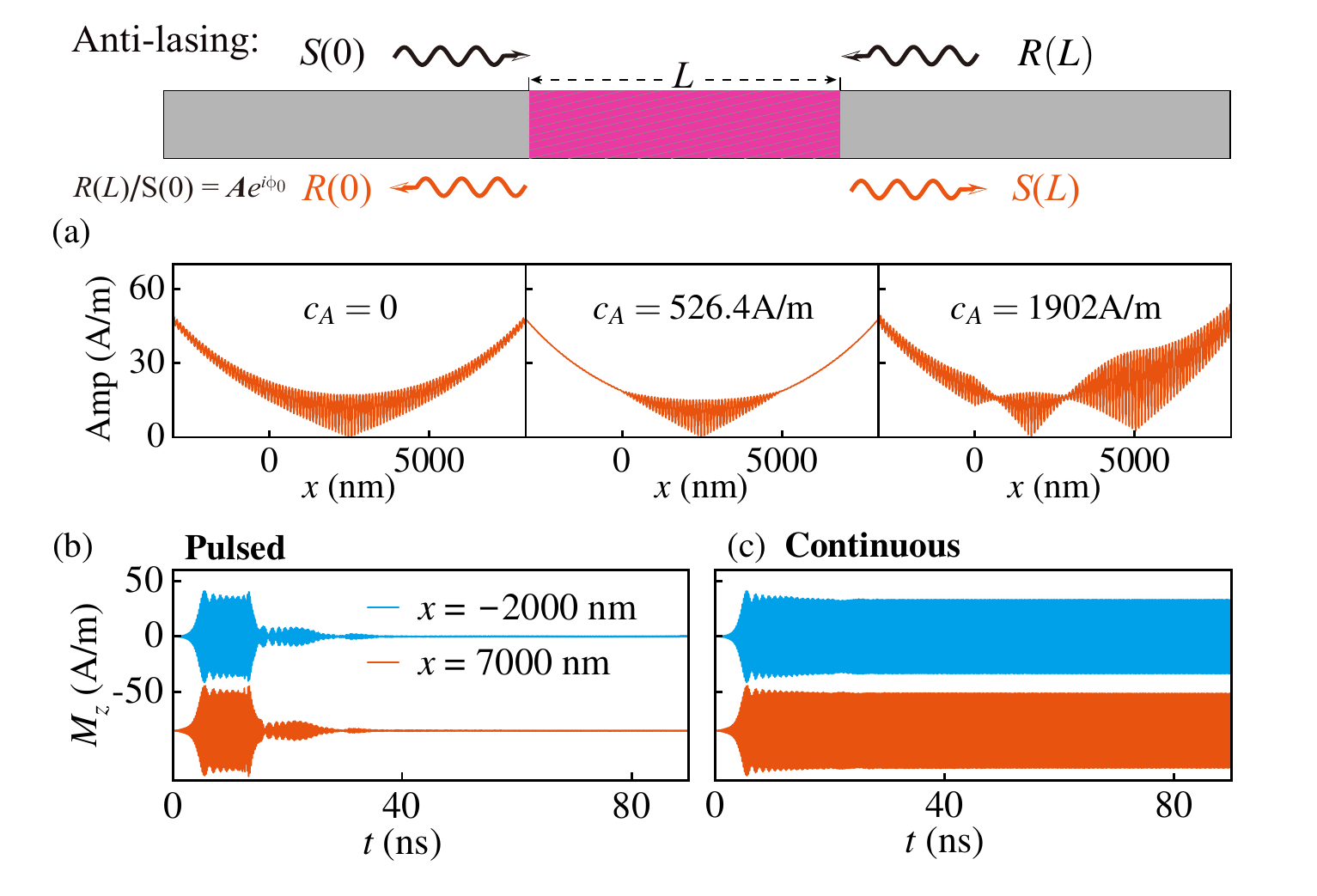}
	\caption{\label{cpa} Top panel is the schematic for magnon anti-lasing with two coherent inputs $R(L) = S(0) A e^{i \phi_0}$. In the micromagnetic simulation with step-type electric current region in $ 0 \le x \le 5000$ nm, we apply two microwave fields $ \vec{h}_1 = h_0 \sin(2 \pi f t) \vec{z} $ (at $x = -4000$ nm) and $ \vec{h}_2 = A h_0 \sin(2 \pi f t + \phi_0) \vec{z} $ (at $x = 9000$ nm) to realize the inputs. Here, $A = 1$ and $ \phi_0 = -1.26 $ are used to approach the absorber condition, $f = 4.701 $ GHz and $h_0 = 1000$ A/m. (a) The spatial profiles of $M_z$ oscillation amplitude under different $c_A$. (b-c) At the critical value $c_A = 1902$ A/m, the time dependent $M_z(t)$ profiles exited by (b) 10 ns pulsed and (c) continuous microwave fields, where the red curves with $|M_z| < 50$ A/m are intentionally offset for clarity.}
\end{figure}

In the numerical simulations, we adopt the step function of $c_J(x)$ [Fig. \ref{model}(b)] to generate the spatially varying gain and loss. The setting is more close to the realistic structure consisting several independent charge-carrying stripes. Simulations are done for a $ 2 {\rm nm} \times 1 {\rm \mu m} \times 20 {\rm nm} $ unit cell covering the total waveguide size of $ 30 {\rm \mu m} \times 1 {\rm \mu m} \times 20 {\rm nm} $, and Eq. (\ref{LLG}) is numerically solved by a finite difference method. In a single period of $ c_J(0 < x \le \Lambda) $, the step function is $c_J(0 < x \le 2\Lambda/5) = c_A,  c_J(\Lambda/2 < x \le 9\Lambda/10) = -c_A$. When its amplitude is $0.825$ times $ c_A$ of sin function, the step function has the same component with the sin function at $k_p$ in the wave vector space. As proved by Figs. \ref{laser}(a,c), for the laser condition $c_A = 0.825 \times 2305 = 1902 $ A/m, the injected magnon pulse (with frequency 4.701 GHz) is sustained for a long time without decaying. Without SOT ($c_A = 0$), the magnon pulse is swiftly  damped. With continuous magnon injection for a long enough duration (say 300 ns) a strong magnon amplification is achieved, see Figs. \ref{laser}(b, e). 
With the large magnon amplitude one enters the magnonic nonlinear regime where two very close frequencies (4.701 GHz and 4.703 GHz) are simultaneously generated, and thus beating with long period is observed, see Fig. \ref{laser}(e). Adopting a slightly smaller $c_A = 1860$ A/m in Fig. \ref{laser}(f), the SOT induced magnon enhancement becomes weaker, and the nonlinear effect is suppressed, where only 4.701 GHz is left. As for a larger $c_A = 1943$ A/m above the critical value, a further larger magnon amplitude is excited, and the main peak appears at 4.718 GHz, see Fig. \ref{laser}(g). In this case, as $\Re[\tau] > 0$, the magnon amplitude is still continuously amplified even under a short pulse excitation and one obtains similar curves (not shown here).

The above $\mathcal{PT}$-symmetric magnon laser has the advantage of selectively amplifying the magnon mode at the Bragg point with quite narrow bandwidth (even in the nonlinear regime), while other magnon modes remain unamplified. If we introduce a pure gain (say $c_J \ge 668$ A/m) on the whole sample to compensate the intrinsic magnon damping at Bragg point, magnon modes with smaller frequencies are all amplified, causing  multi-mode excitation. Also, the local magnetization is soon reversed and the magnons are damped. To sustain  the magnetization oscillation and amplify magnons, a tilt angle between the electric polarization and equilibrium magnetization is usually required, while the mode selection is still challenging \cite{Haidar2019, Fulara2019, Houssameddine2007, Kaka2005}. Here, the $\mathcal{PT}$-symmetric magnon laser avoid such limitations.

Adding  other coherent magnon injection to the same structure, simulation results in Fig. \ref{cpa} evidence   anti-lasing under a different $c_A$. Without SOT ($c_A = 0$), the interference between two coherent magnon injections causes  amplitude fluctuation in Fig. \ref{cpa}(a), where its spatial profile is affected by the magnon decaying. At the anti-lasing condition $c_A = 0.825 \times 638 = 526.4 $ A/m, the injected magnons are absorbed in the region of $ 0 \le x \le 5000$ nm, and the amplitude fluctuation outside the electric current region is completely erased. Noteworthy, as the step function is not exactly same to sin function, the phase $\phi_0 = -1.26$ is different from analytical estimation while the amplitude ratio $A = 1.0$ still follows $|M_{21}| = 1$. Switching to the laser condition $c_A = 1902 $ A/m, two persisting coherent magnon injections totally suppress the laser amplification, while there are still clear reflections. Such observation agrees with above theoretical estimations, where $U = 0.67$ is found. As the analytical $M_{21}$ here is identical to the anti-lasing condition, $A = 1$ and $ \phi_0 = -1.26 $ are still applicable. Furthermore, Figs. \ref{cpa}(b-c) show the time-dependent results for the magnon amplitude under pulsed and continuous excitation, and indeed the magnon amplification is completely suppressed. We also considered different $A$ and $\phi_0$, and except for the above condition the magnon laser function can be fully resumed.

\begin{figure}[tbp]
	\includegraphics[width=0.5\textwidth]{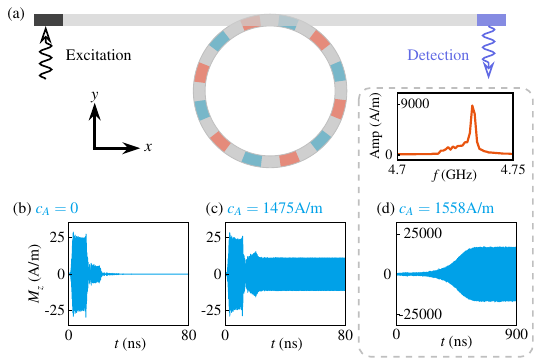}
	\caption{\label{circle} (a) Schematic for the magnetic ring structure (gray region) with periodically varying SOT induced gain (blue regions) and loss (red regions). The ring is coupled to a planar waveguide. Magnons are injected at the left side of waveguide, amplified in the ring laser, and detected at the right side in the waveguide. For (b) $c_A = 0$ and (c) $c_A = 1475$ A/m, time dependent $M_z$ at $x = 400$ nm in the waveguide. Magnons are injected at $x = - 600$ nm by applying $ h_0 \sin(2 \pi f t) \vec{z} $ lasting 10 ns with $h_0 = 1000$A/m and $f = 4.73$ GHz. (d) At $c_A = 1558$ A/m, time dependent $M_z$ at $x = 400$ nm under finite temperature $T = 1$ K, with the above frequency spectrum. Results are obtained from micromagnetic simulations.}
\end{figure}

To tune the magnon lasing and anti-lasing frequency, a different periodic length $\Lambda$ can be set to change the Bragg point position. As an example, with $\Lambda = 500$ nm and $ L = 1000$ nm, the laser absorber for magnon frequency 3.56 GHz is achieved (details see the Supplemental Material). Noteworthy, for such setting, only four separated charge-carrying stripes are enough, validating the effectiveness of the simpler structure for the magnon laser absorber. Alternatively, we suggest to change the Bragg period by rearranging the electric current density profile in the charge-carrying stripes or adjusting the external magnetic field, which highlights the electrical or magnetic tunability of the magnon laser frequency. In addition, the lasing and anti-lasing mechanism is applied to antiferromagnets rendering possible the selective excitation of one of two oppositely polarized magnon modes in the THz frequency range (Supplemental Material).

The mangon laser mechanism also applies for a magnetic ring. As demonstrated by Fig. \ref{circle}(a), to generate the magnonic laser, the ring is exposed to  SOT induced change between gain and loss with a period corresponding to an arc length of $\pi/25$ radians. In the single period, the step function $c_J(0 < \psi \le 2\pi/125) = c_A,  c_J(\pi/50 < \psi \le 9\pi/250) = -c_A$ is used. We set a ring with an average radius of 780 nm, width 40 nm and thickness 40 nm. The period length $\Lambda$ is approximately 98 nm, and at the Bragg point the magnon frequency is 4.73 GHz. The magnons are injected in a planar waveguide and coupled to the ring via the interlayer coupling over $-250 {\rm nm} < x < 250 {\rm nm}$. In Figs. \ref{circle}(b-c), the simulation results for injected magnon pulses under $c_A = 0$ and $c_A = 1475$ A/m are compared, and the laser condition induced unattenuated magnons are obvious. We also validate the amplification of thermal magnons under a finite temperature. The uniform temperature $T = 1$ K is introduced via the thermal random magnetic field with the white-noise correlation function  $ \langle h_i(\vec{r},t) h_j (\vec{r}', t) = \frac{2k_{\mathrm{B}} T \alpha}{\gamma M_{\mathrm{s}} V} \delta_{ij} \delta(\vec{r} - \vec{r}') \delta(t-t') \rangle $, where $k_B$ is the Boltzmann constant and $V$ is the volume. In Fig. \ref{circle}(d), adopting a slightly larger $c_A = 1558$ A/m, thermal magnons near the Bragg point are selectively amplified.

To conclude,  a $\mathcal{PT}$-symmetry-based  setup for  lasing and anti-lasing of magnons is proposed. The device is solely driven by dc currents that generate a controlled, periodically modulated gain–loss landscape implementing non-Hermitian magnon dynamics. Analytical modeling uncovers the essential physics complemented with full numerical simulations that confirm  single magnon mode amplification at the Bragg point,  absorption for coherent inputs at the lasing frequency, and continuous electrical tunability of emission frequency and output level. Systematic studies over current density, geometric length, and damping provide practical guidelines for device realization. The operating principle naturally extends to planar waveguide and ring geometries, where the waveguide enables on-chip integration and the ring supports stable single-mode operation. The active region also amplifies thermal magnons, pointing to low-bias startup and potential noise-shaping strategies. These results provide a clear  pathway toward  electrically programmable magnonic sources, sinks, and amplifiers within a single, compact component.

\textit{Acknowledgments}—This work was supported by the National Key R\&D Program (Grant No. 2024YFE0102400), the National Natural Science Foundation of China (Grants No. T2495212,  No. 12174452, No. 12274469, No. 12074437, No. 11935006, No. 12421005 and No. 12565001), DFG under project nr. 465098690, the Hunan Major Sci-Tech Program (Grant No. 2023ZJ1010), and the Natural Science Foundation of Hunan Province of China (Grants No. 2025JJ20005).

\end{document}